# Thermodynamic parameters of the η′ phase in an AlZnMgCu alloy synthesized by mechanical alloying

**María del V. Valera M.[1] and Ney J. Luiggi A.[2,*]**
Grupo Física de Metales, Dpto. de Física, Núcleo de Sucre-Universidad de Oriente.
Cumaná, Venezuela.
[1] E-mail: mvalera05@gmail.com.  [2] E-mail: nluiggi51@gmail.com.  (*) Correspondance author

**Abstract**

We synthezised an Al-Zn-Mg-Cu alloy through mechanical alloying and, using X-ray diffraction (XRD), identified the formation of the η′ phase after 40 hours of grinding. Using reaction-free isoconversion theory, we determined that this phase exhibits two different behaviours depending on the heating rate ($\beta$): the η′ phase at low $\beta$ and the η phase at high $\beta$. The activation energy values at low $\beta$ are consistent with the diffusion energies of copper, zinc and magnesium in aluminium. Significant qualitative and quantitative differences in the thermodynamic barriers ($\Delta H$, $\Delta G$ and $\Delta S$) are observed for $\beta$ values above or below 20 °C/min.

1. **Introducción**

Mechanical alloying is a rapidly growing method of synthesizing new materials due to its versatility and ability to create nanometric structures, which are currently very popular in materials science [1-5]. Cyclic mechanical deformation of powders enables elements, even immiscible ones, to be welded together to form initially metastable phases that become stable through appropriate heat treatment [6–11].

This work focuses on grinding Mg, Zn, Cu and Al powders in the same proportions found in the AA7075 alloy. A sequence of phases has been identified in this alloy, including GP zones, the η' phase, the η phase and complex Cu phases at higher temperatures. The microstructure and population of each of these phases, which are controlled by heat treatments, are responsible for the excellent mechanical properties of this alloy [12-17]. Therefore, it is important to understand the kinetics involved in forming these phases, as well as the thermodynamic barriers, such as the activation energy, enthalpy variation, variation in Gibbs free energy and variation in entropy, generated by the transition to the new phase.

The kinetics of solid-solid reactions have been extensively studied within reaction rate theory, which is entirely defined by kinetic parameters and the kinetic function that governs the reaction. To determine these variables, kinetic models with specific reaction mechanisms must be used where this function is preset. Isoconversion models that are free of a kinetic function fulfil this objective since, when different heating rates are applied and the conversion extension is set, the kinetic function assumes the same value, leaving the temperature and heating rate as dependent variables. As an analysis technique, isoconversion has transcended the boundaries of kinetic studies and become a versatile tool, with the work of Sergey Vyazovkin *et al.* [18-22] being particularly notable.

In this work, we synthesized Cu, Zn, Mg, and Zn powders by mechanical alloying, demonstrated the formation of the η´ phase, and determined the thermodynamic barriers of the reaction using kinetic function-independent isoconversion models, namely Ozawa-Flynn-Wall (OFW), Friedman (F), and Kissinger-Akira-Sunose (KAS) kinetic function, as well as the modified Kissinger and Friedman models, which are non-isoconversion and mainly dependent on the peak reaction temperature.

## 2. Experimental methodology

### 2.1 Experimental aspects

High-purity commercial powders were used to prepare the samples: Al (99.9%), Mg (98%), Zn (98%) and Cu (99.9%). These were mixed in an argon environment in a Fritsch PULVERISETTE 7 Premium Line planetary mill with a ball-to-powder mass ratio of 40:5, using proportions equivalent to those used in preparing the AA7075 alloy (5.6 wt.% Zn, 2.5 wt.% Mg and 1.6 wt.% Cu). The grinding times were: 1, 5, 7, 10, 20, 30, 40 and 50 hours.

The progress of the alloy was checked by XRD at the end of each grinding period. This was performed using a PANalytical X-ray diffractometer (model X'Pert3 Powder) with Cu–K$\alpha$ radiation ($\lambda$ = 0.1542 nm). Once the samples had been consolidated and the presence of the different phases (including the η′ phase) had been confirmed, they were subjected to a calorimetric study using a Perkin–Elmer TAC 7/DX DSC 7 calorimeter. The samples were placed in aluminium vials at a mass of 30 mg each; the reference was a vial containing pure aluminium. The study was performed at heating rates of 5, 10, 20, 50 and 75 °C/min, ranging from room temperature to 600 °C.

### 2.2 Theoretical aspects

The calorimetric signal reflected by the thermograms obtained from DSC measurements provides information on all the processes involved in the transition reaction. Its theoretical study within the framework of reaction rate theory has been widely reviewed in the literature. According to this theory, the change in the extent of conversion $\alpha$ over time or temperature depends on the reaction constant K(T) and a kinetic function F($\alpha$), which involves the mechanisms that generate the reaction.

$$\frac{d\alpha}{dt} = K(T)F(\alpha) \qquad (1)$$

For thermally activated processes, the Arrhenius equation establishes the exponential dependence of K(T) on (-Q/RT), where Q is the minimum energy required to activate the reaction. When analyzed in both its differential and integral forms, equation (1) generates the various expressions employed in the literature for studying reactions in the solid state.

The derivative of the reaction rate leads to

$$\left(\frac{d^2\alpha}{dt^2}\right) = \left[\left(\frac{Q \cdot \beta}{R \cdot T^2}\right) + K(T)\frac{dF}{d\alpha}\right] \cdot \frac{d\alpha}{dt} \qquad (2)$$

Where the heating rate $\beta = dT/dt$ is constant, a linear dependence is defined between T and t.

Relationship (2) forms the basis of Kissinger's model [23], to which certain conditions are applied: Firstly, the rate of change of the reaction velocity shows a minimum, $(d^2\alpha)/(dt^2) = 0$, which makes the expression in brackets in relationship (2) equal to zero. This occurs at a temperature $T = T_P$. Secondly, although F is a function of the conversion extent, its derivative remains constant and equal to unity in Kissinger's model, which presets this kinetic function to a first-order reaction. This leads to the following expression:

$$\ln\left(\frac{\beta}{T_P^2}\right) = \ln\left(-K_0 \cdot \frac{R}{Q}\left(\frac{dF}{d\alpha}\right)_P\right) - \frac{Q}{RT_P} \tag{3}$$

Including $\left(\frac{dF}{d\alpha}\right)_P$ explicitly in relation (3) extends the validity of Kissinger's model to kinetic functions other than first order, thus making it dependent on the selected kinetic model. Note also that $\left(\frac{dF}{d\alpha}\right)_P < 0$ must be satisfied. Luiggi [24] also addresses equation (1) through the series expansion of the exponential integral, reporting the following expressions:

$$\ln\left(\frac{d\alpha}{dT}\right) - \ln F(\alpha) = \ln\left(\frac{K_0}{\beta}\right) - \frac{Q}{RT} + \ln\left[\frac{R}{Q}\left(\sum_{M=1}^{\infty}(-1)^M(M+1)!\left(\frac{RT}{Q}\right)^M\left(1 + M + \frac{Q}{RT}\right)\right)\right] \tag{4}$$

$$\ln\left(\frac{f(\alpha)}{K_0} - \frac{K_0}{R\beta}C(T_0)\right) = \ln\left(\frac{T^2}{\beta}\right) - \frac{Q}{RT} + \ln\left[\frac{R}{Q}\left(\sum_{M=1}^{\infty}(-1)^M(M+1)!\left(\frac{RT}{Q}\right)^M\right)\right] \tag{5}$$

Where $f(\alpha) = \int \frac{d\alpha}{F(\alpha)} = \int K(T(t))dt = I$ y $C(T_0)$ represents the contribution of the initial temperature to the integral I [13]. According to equation (1), the expression in brackets in equation (5) is equal to unity. This allows the number of terms in the series expansion of I to be fixed. However, without significant loss of accuracy, the terms to be used in the summations of expressions (4) and (5) can be restricted to the first two terms [24]. The study models used in this work are summarized in Table 1.

Table 1. Methods of analysis used in this study.

| Method | Relationship to be plotted | Cut | Slope | Reference |
|---|---|---|---|---|
| Kissinger (K) | $ln\left(\frac{\beta}{T_P^2}\right)$ vs $\frac{1}{T_P}$ | $ln\left(\frac{K_0 R}{Q}\right)$ | -Q/R | [23] |
| Generalized Kissinger (GK) | $ln\left(\frac{\beta}{T_P^2}\right)$ vs $\frac{1}{T_P}$ | $ln\left(-K_0 \frac{R}{Q}\left(\frac{dF}{d\alpha}\right)_P\right)$ | -Q/R | * |
| Modified Friedman (FM) | $ln\left[\beta\left(\frac{d\alpha}{dT}\right)_P\right]$ vs $\frac{1}{T_P}$ | $ln(K_0 F(\alpha_P))$ | -Q/R | [25] |
| Ozawa-Flynn-Wall (OFW) | $ln(\beta)$ vs $\frac{1}{T_\alpha}$ | $ln\left(\frac{K_0 F(\alpha)}{(d\alpha/dT)_\alpha}\right)$ | -1.052 Q/R | [26,27] |
| Friedman (F) | $ln\left[\beta\left(\frac{d\alpha}{dT}\right)\right]$ vs $\frac{1}{T_\alpha}$ | $ln(K_0 F(\alpha))$ | -Q/R | [28] |
| Kissinger-Akira-Sunose (KAS) | $ln\left(\frac{\beta}{T_\alpha^2}\right)$ vs $\frac{1}{T_\alpha}$ | $ln\left(-K_0 \frac{R}{Q}\left(\frac{dF}{d\alpha}\right)_\alpha\right)$ | -Q/R | [29] |

Assuming that phase transitions occur through nucleation and growth, the nucleation rate J for first-order nucleation is defined as [20, 30–32].

$$J = K_0 \exp\left(-\frac{Q_\alpha}{RT}\right)\exp\left(-\frac{\Delta G}{RT}\right) \qquad (6)$$

$K_0$ represents the Arrhenius pre-factor, while ΔG represents the thermodynamic energy barrier for nucleation. This is defined by the competition between interfacial and volumetric energies during the development of the new phase. The parameters involved in relation (8) represent the thermodynamic barriers for the initiation of random nucleation and obey the following expressions [31, 32].

$$K_0 = \frac{\beta \cdot Q_\alpha \exp\left(\frac{Q_\alpha}{R \cdot T_P}\right)}{(R \cdot T_P^2)} \qquad (7)$$

$$\Delta S = \frac{(\Delta H - \Delta G)}{T_p} \tag{8}$$

$$\Delta H = Q_\alpha - R \cdot T_\alpha \tag{9}$$

$$\Delta G = Q_\alpha + R \cdot T_P \cdot \ln\left(\frac{k \cdot T_P}{h \cdot K_0}\right) \tag{10}$$

These expressions are completely defined with knowledge of the activation energy Q and the pre-factor $K_0$. The activation energy, Q, is evaluated from the slope of the experimental DSC data using each of the methods proposed in Table 1. However, with the exception of the K method, the cut exhibited in Table 1 is contingent upon the kinetic function F(α), which would necessitate the utilization of the kinetic functions documented in the extant literature for this purpose [25]. In our case, we evaluate this pre-factor from equation (3) for first-order reactions, which gives us expression (7), thus defining all thermodynamic barriers.

3. **Results and discussion**

Our DSC measurements, which follow the precipitation sequence GP zones → η′ phase → η phase, enable us to isolate the phase under study between 60°C and 275°C. Figure 1 shows the respective heat flow for different values of the heating ratio β, reflecting an exothermic reaction whose maximum shifts towards higher values as β increases. This behaviour is typical of reactions generated by the diffusion of alloy elements. The respective peak values are in ascending order of β: 127.5, 159.5, 177.42, 197.66 and 214.15 °C.

From Figure 1 we access the value of the conversion extension, $\alpha = \int_{T_0}^{T} H\, dT \, / \int_{T_0}^{T_F} H\, dT$, and its derivative with respect to temperature..

Applying the models in Table 1 generates Arrhenius-type graphs, shown in Figure 2. For each model, it can be seen that there are two sets of data that align in different directions. The first set corresponds to values of β ≤ 20°C/min and has a steeper slope, while the second set corresponds to higher heating rates and has a shallower slope. This demonstrates the occurrence of different reaction mechanisms at high and low β values.

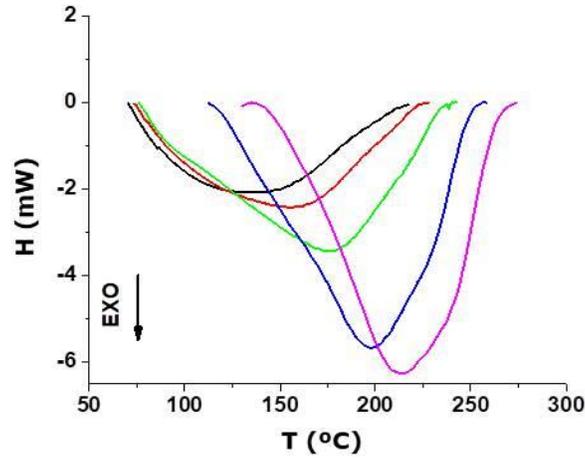

Figure 1. Thermogram corresponding to the formation of the η' phase in the AlZnMgCu alloy at different heating rates. ▬ 5°C/min, ▬ 10°C/min, ▬ 20°C/min, ▬ 50°C/min, ▬ 75°C/min.

Figure 2 shows the apparent activation energy as a function of the transformed fraction for each model, determined from the slope of the aligned points. Figure 3 shows these results and illustrates that the dependence of Q on α is qualitatively similar in the KAS, F and OFW models, with more pronounced quantitative differences at the beginning and end of the reaction. Similarly, significant differences in Q are observed depending on the value of β; the reaction occurs more easily at high values of β in terms of energy. In previous analyses [4], using methods where Q is independent of α, .all results were attributed to the η′ phase

The Q values in the modified Kissinger and Friedman models are not isoconversional. They are calculated based on the temperature values at the peak of maximum heat flow and are more consistent with the results obtained for high β values.

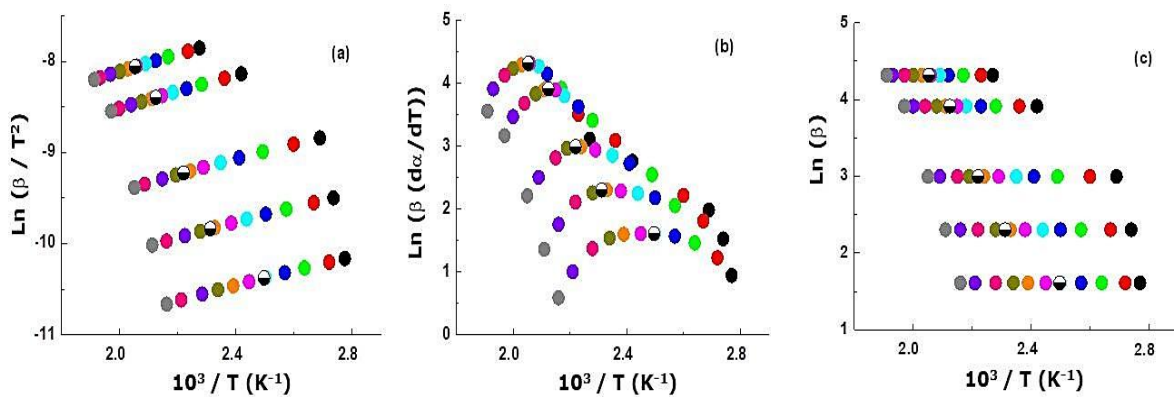

Figure 2: Arrhenius plots are used to determine the apparent activation energy of the η′ phase. The coloured balls correspond to the conversion range from: ● α=0.05 to ● α=0.95, with steps of 0.05. (a) KAS model: (b) Friedman model. (b) OFW model. ◕ The Kissinger model is shown in (a). ◕ The modified Friedman model is shown in (b) and (c).

Figure 3 confirms the occurrence of two different reactions or mechanisms in the formation and consolidation of the η′ phase, or the formation of different phases at low and high heating rates. At low β, the process is controlled by the diffusion of Cu, Zn and Mg, whose activation energies for diffusion are consistent with those shown in the figure for low and high α, while in the intermediate region, our results are consistent with those obtained by García-Cordovilla and Louis [33] in an Al-Zn-Mg-Cu alloy subjected to retrograde treatment. However, low Q values for high heating rates could affect the diffusion of grain boundaries, which has a lower activation energy, and cause the η phase to form prematurely. According to DeIasi et al. [34], the activation energy of the η phase fluctuates between 50 and 66 kJ/mol.

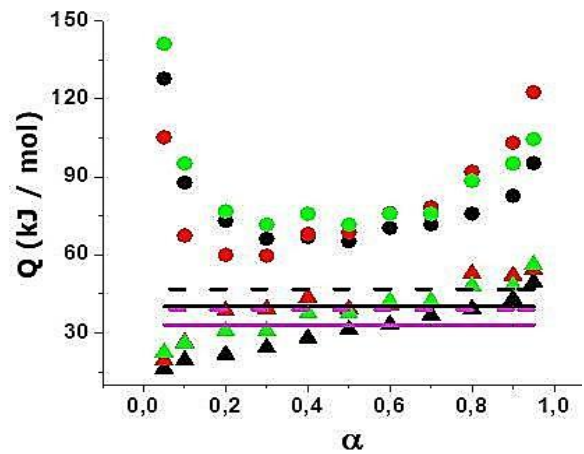

Figure 3: Apparent activation energy versus transformed fraction, deduced from Figure 2 for each of the models defined in Table 1: ● ▲ KAS model, ● ▲ Friedman model and ● ▲ OFW model. Balls correspond to β ≤ 20°C/min. Triangles correspond to β > 20°C/min. ——— K model (β ≤ 20°C/min). ——— FM model (β ≤ 20°C/min), ------K model (β > 20°C/min). ----- FM model (β > 20°C/min):

The evaluation of thermodynamic barriers, as defined by equations (8) to (10), requires certain considerations. 1. $K_0$ is evaluated from expression (7), which is valid for first-order reactions and framed within temperature-dependent maximum flow models. 2. As the values of Q(α) are derived from various β values, the value of α(T) is taken as the average curve of the transformed fractions for each β value, as illustrated in Figure 4.

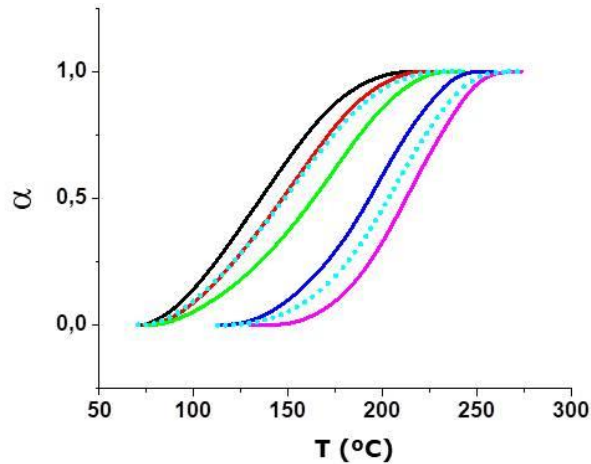

Figure 4. Transformed fraction as a function of temperature for different values of β, showing in dashed lines (------, ------) the average value of α(T) for low and high β. ▬ 5°C/min, ▬ 10°C/min, ▬ 20°C/min, ▬ 50°C/min, ▬ 75°C/min.

Based on these considerations, Figure 5 illustrates the evolution of $K_0$, $\Delta H$, $\Delta G$ and $\Delta S$ of the phases derived from Figure 3. As can be seen in this figure, the Arrhenius pre-factor, which is associated with the frequency of atomic collisions required for the reaction to occur, exhibits qualitative behaviour determined by the activation energy. Values are much higher for $\beta \leq 20$°C/min than for higher β values.

The near-zero value of $K_0$ for the η phase at temperatures below 220 °C suggests that this phase has not yet formed. In general, thermodynamic barriers exhibit similar behaviour at temperatures above 150°C. The positive variation of $\Delta H$ always indicates that neither reaction occurs spontaneously, with phase η' being less stable than phase η, as reflected by the higher energy required to activate phase η'. This behaviour is confirmed by the variation in Gibbs free energy. The different values of $\Delta G$ for the two phases indicate different internal energies. This reinforces our hypothesis that different phases occur at different heating rates. Regarding the entropy variation, negative values in the η phase indicate phase condensation due to an increase in order within the system. In contrast, the η' phase exhibits positive values of $\Delta S$ at low and high temperatures, which is associated with the dissolution of the previous phase (GP Zones in our case) or the dissolution of the η' phase itself.

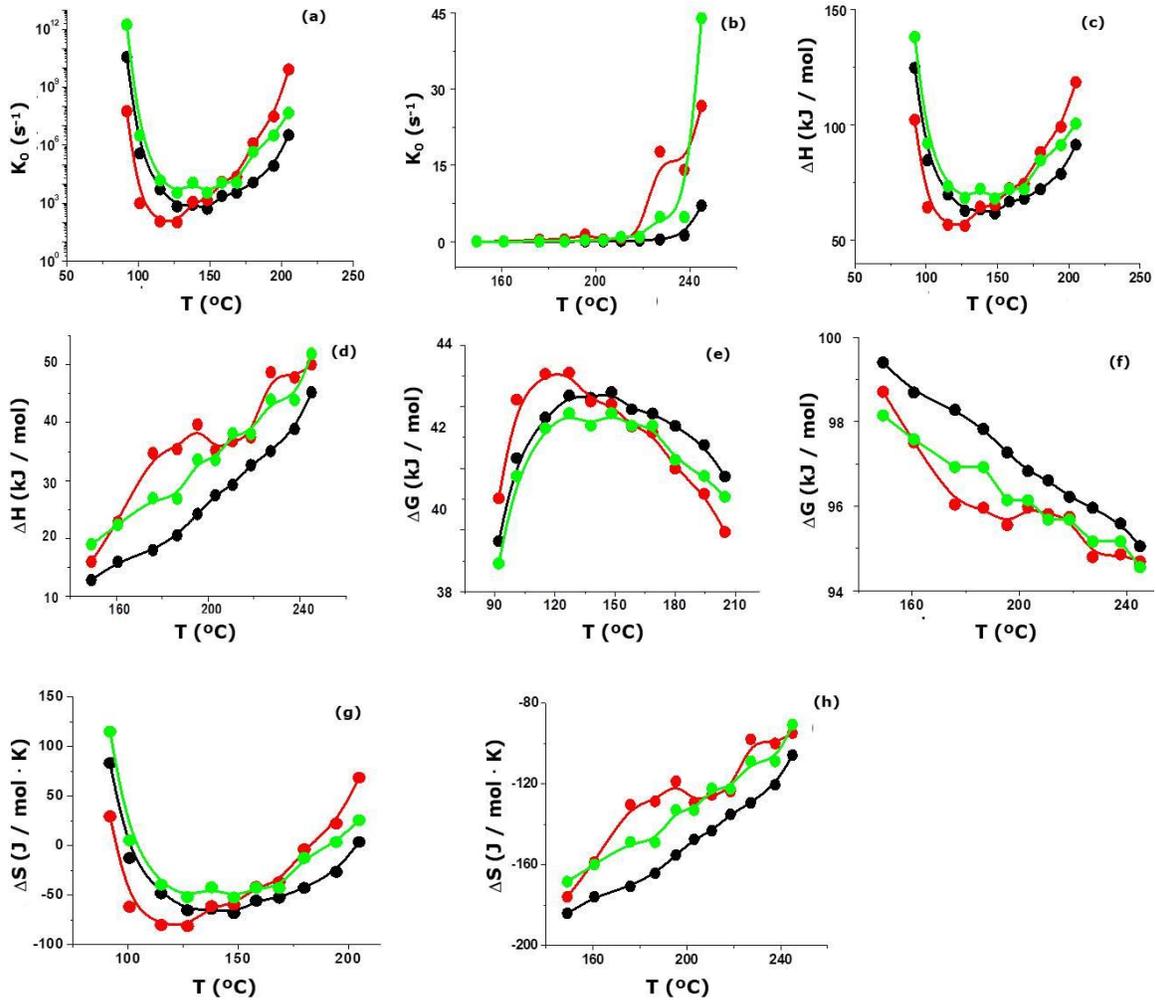

Figure 5. Pre-factor and thermodynamic barriers deduced from the isoconversion model for the evolution of phase η' (β ≤ 20°C/min). ● KAS model. ● Friedman model. ● OFW model.

Overall, these results demonstrate the impact of the heating rate on the formation of the η' and η phases in our Al-Zn-Mg-Cu mechanical alloy.

Conclusions:

We conducted a DSC study on the formation of the η' phase in an Al-Zn-Mg-Cu alloy prepared by mechanical alloying. The study revealed the following:

- ❖ The formation of the η' phase occurs through the diffusion of alloying elements.
- ❖ At high heating rates, a phase with different thermodynamic parameters to those of the η' phase is evident. We associate this phase with the η phase due to the calculated activation energy values.

- ❖ Three isoconversional models, free of kinetic function, were used to determine the thermodynamic barriers of the process. The resulting values were consistent with each other and comparable to those reported in the literature.
- ❖ Modified Kissinger or Friedman models generate activation energy values that are lower than those reported in the literature for the η′ phase.
- ❖ Values of ΔH, ΔG and ΔS confirm formation of the η′ phase at low values of β and formation of the η phase at values of β above 20 °C/min.